\documentclass[12pt]{article}

\usepackage{times}
\usepackage{epsf}
\usepackage{epsfig}

\newcommand{\isotoday}{\space\number\year-{\ifcase\month\or
   Jan\or Feb\or Mar\or Apr\or May\or Jun\or
   Jul\or Aug\or Sep\or Oct\or Nov\or Dec\fi}-{\ifnum\day<10
   0\number\day \else \number\day \fi}}
\newcommand{\simgt}{\raisebox{0.3ex}{$> \kern -1em {\lower 1ex \hbox{$\sim$}}$\thinspace}}
\newcommand{\simlt}{\raisebox{0.3ex}{$< \kern -1em {\lower 1ex \hbox{$\sim$}}$\thinspace}}
\def\bib#1{\par\noindent\hangindent=1cm\hangafter=1 #1 \bigbreak}

\addtocounter{secnumdepth}{2}
 
\begin{document}

\baselineskip=1.1\normalbaselineskip

\sloppy

\centerline{\Large\bf Investigations of the Formation and Evolution of  Planetary}
\vspace*{2truemm}
\centerline{\Large\bf Systems}

\
                                                                                
\centerline {\bf Alwyn Wootten, Bryan Butler, Antonio Hales, Stuartt Corder,
\footnote{National Radio Astronomy Observatory}} 
\centerline{ \bf Robert Brown\footnote{National Astronomy and Ionosphere Center} \& David Wilner\footnote{ Harvard-Smithsonian Center for Astrophysics}} 

{\bf Abstract.}
Stars and planets are the fundamental objects of the Universe.  Their formation processes, though related, may differ in important ways.  Stars almost certainly form from gravitational collapse and probably have formed this way since the first stars lit the skies.  Although it is possible that planets form in this way also, processes involving accretion in a circumstellar disk have been favored.  High fidelity high resolution images may resolve the question; both processes may occur in some mass ranges.  The questions to be answered in the next decade include:

By what process do planets form, and how does the mode of formation determine the character of planetary systems?

What is the distribution of masses of planets?  In what manner does the metallicity of the parent star influence the character of its planetary system?

In this paper we discuss the observations of planetary systems from birth to maturity, with an emphasis on observations longward of 100 $\mu$m which may illuminate the character of their formation and evolution.  Advantages of this spectral region include lower opacity, availability of extremely high resolution to reach planet formation scales and to perform precision astrometry and high sensitivity to thermal emission.


{\bf Introduction.}
Planets are cool bodies and emit through reflected light from their host star and through their own blackbody radiation.  Since their mass is a fraction of that of their parent stars, direct radiation from planets has been difficult to detect; most of our information on extrasolar planets is derived from indirect observation, usually at optical and infrared wavelengths.   Planets form either as stars do, through gravitational collapse, or through accretion of material in a remnant disk.  During early phases under the latter scenario, the large luminous regions in which accretion occurs may be directly observable.  As the condensations grow and sweep up material from their natal disks, structures in the disks can signal the presence of embedded planets even when the condensations themselves are not visible.  A wobble in the position of the central star is induced by the planetary retinue which may be detected through precision astrometry.  Finally, as interactions between bodies distribute debris in the disks, gaps due to planetary sweepup of material may provide additional clues as to the presence of planets.  Wavelengths longer than 100 $\mu$m can offer complementary information on the character of planets and planetary systems.  Here we explore some observable effects.

\begin{figure}[ht]
\begin{center}
\includegraphics[scale=0.4, angle=-0] {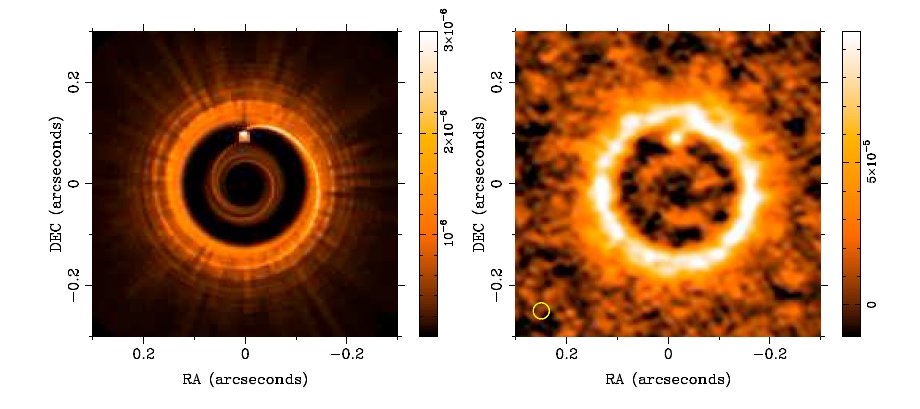}
\caption{Left-hand panel: Model image of a proto-Jupiter embedded in a 0.02 M$_{Sun}$
protoplanetary disk located 50 pc away from the Sun (courtesy S. Wolf, from
Wolf \& DÕAngelo (2005)). The orbital radius is 5 AU, and the mass of the central star
(not included in this image) is 0.5 M$_{Sun}$. The original image has been scaled from 900 to
345 GHz assuming S$\propto \nu^2$. Right-hand panel: A 345 GHz, 8 hour simulated observation of
S. WolfÕs model, using one of ALMAÕs extended configurations, with a 25$\times$20 mas FWHM beam (using Briggs weighting and maximum baseline = 8 km, represented by the ellipse in the lower-left
corner).}\label{fig1}
\end{center}
\end{figure}

{\bf Imaging Very Young Systems.}
Observations indicate that nearly all
young low mass stars are born surrounded by 
disks of molecular gas and dust.
Such disks appear to be a natural feature of the process of star and 
solar system formation.  The observations show that the properties of 
the disks are similar to those predicted to have existed in the nebula 
from which our solar system is presumed to have formed.  These disks 
may therefore provide tests of the theories of how our solar system 
formed, and how other systems may form in general.

As the youngest protostars begin to form, surrounding material hides
them from view at all but the longest wavelengths.  The reservoir of 
cold material, still to be accreted by the protostar, contributes 
appreciably to the total luminosity of the system.  One of the most 
effective means of locating young protostars then is to detect that 
long wavelength emission and to contrast long wavelength luminosity to 
total luminosity.  Since luminosity is measureable, and to first order 
proportional to mass, this ratio tracks the ratio of circumstellar to 
stellar mass, which decreases with age over the first few hundred 
thousand years of evolution.  Finally, the leftover material swirling 
in a disk may become incorporated into planets, during the first 
several million years of a star's life.  

It seems likely that the physical conditions in these very young disks determine the characteristics of the planetary systems which will form.  Currently, the resolution and sensitivity to characterize these physical conditions on the scales of planetary orbits within very young disks is lacking.  Most systems with stars young enough to harbor disks in their dustiest and gassiest states lie at distances of just over 100pc from us, so that even the Solar System subtends less than one arcsecond.  

Young stellar systems emit continuum radiation from the central stars and from dust at a wide range of temperatures in circumstellar disks.  With sufficient wavelength coverage and sensitivity the  spectral energy distributions (SED)s of these young star-forming regions can be decomposed to determine which of them possess disks and to begin to constrain the character of those disks.  Statistics on the fraction of young stars with disks, and on the overall character of those disks can be improved with large format detectors on the large telescopes expected to become available in the next decade; these will help constrain theories of their formation.  Distinctions between different types of low mass stars and their formation mechanisms may be made possible (i.e., between brown dwarfs, various types of low mass stars, and even with sufficient sensitivity, whether planets must form through processes secondary to star formation).

The world's leading optical observatories are building the first generation of high contrast
ÒextremeÓ adaptive optics  systems, which deliver imaging performance at
scales $>$ 0.1 arcseconds.  As the new generations of 30m class optical telescopes are built and outfitted with a new generation of adaptive optics systems, imaging at scales of ~20 milliarcseconds will become available in the infrared.  These systems will delve into the outer regions of less-embedded solar systems by detecting the radiation from superJovian planets but it will be difficult to separate starlight from scattered lightÊin the disks.  Complementary images may be made in the submillimeter, where resolutions of $\sim$ 10 milliarcseconds will be realizable and should  reveal extremely young Jovian mass planets still in their accretion phases (Figure 1, Wolf and D'Angelo 2005).    These images would provide roughly 60 ``pixels'' 
across such a disk in the Taurus molecular cloud (at 140  parsecs distance).  Even at the distance of the Orion cloud (at 400 pc), about 20 pixels would be obtained.  Images with  such resolution could be used to characterize the disks superbly.    These high resolution images will enable a number of studies not possible with current instrumentation.  Dust imaging determines disk size, physical temperature and dust density as a function of distance from the central star.  Multi-frequency observations will be used to constrain the value 
of $\beta$ (the dust grain opacity exponent) quite accurately.  The variation of $\beta$ with wavelength is diagnostic of grain size and therefore of the early stage of planet formation.
The  delineation of these parameters from system to system will provide insight into the planetary formation process.  By observing many systems of different age, evolution of these properties will be investigated.

Submillimeter line emission provides an important mechanism for gas cooling and characterization of the emission with high resolution and sensitivity probes physical conditions.  Heterodyne spectroscopy offers outstanding velocity resolution, probing the dynamics of the young disk.
As gas and dust from the surrounding disk is accreted onto these cores, 
some become luminous enough to be detected at submillimeter wavelengths.  As these 
condensations continue to grow, they are expected to clear gaps or 
inner holes in the protoplanetary disks.  These gaps and holes are not 
only predicted theoretically, but are
inferred from the SED's of young systems 
and the inner holes have actually been observed in several systems 
e.g. TW Hya, GM Aur and LkCa 15.

Evolution of the disks appears to occur rapidly.  While some dust appears to remain through the evolution from protoplanetary (ages of a few 10$^5$ years) to transitional (ages to 10$^7$ years), gas appears to be rapidly depleted.  The history of that depletion lends insight into the evolutionary processes shaping the planetary system.  Measurements of the gas lend insight into the changing physical conditions within the disk in those earliest times.  In the accretion picture of planet formation, gas must remain through the long-lived assembly stages, while in the gravitational instability model of planet formation gas disappears relatively quickly.  Gravitational instability, if it exists, may also vary in importance with time causing a blending of these two approaches.

Images of molecular line emission would be a valuable tool in determining
the dynamics of these disks and may provide a means of direct detection
of signatures of gravitational instability.  Particularly, it could be
diagnosed whether these disks are currently in Keplerian rotation, or are
undergoing gravitational infall, or both, and whether this changes as a
function of age in these systems. 

Images of molecular line emission could also be used to 
answer many questions about the chemistry in these disks, including: is there is a change from 
kinetically controlled ``interstellar'' chemistry in the 
outer part of the disk, to equilibrium dominated 
``nebular'' chemistry in the inner part?  Why are 
molecular species in these disks depleted relative to 
their usual interstellar abundances?  How does grain 
boundary chemistry affect the overall chemical structure 
of the disk?  Chemical gradients of a great number of 
molecular species will be detectable, should they exist.  CO is a dominant coolant, emitting throughout the millimeter and submillimeter range from these warm environments.  Water, also a dominant coolant,  emits in a number of lines.  Images of disks, where gas temperatures are typically 100-300K,  might be made in the low energy ($\sim$200K) 183 GHz line of water, something not possible from Earth except at high very dry sites.  

%



{\bf Direct Detection}  The possibility of direct detection of planets around other stars 
is an exciting one, and we explore that possibility here for millimeter/submillimeter wavelengths.  
By direct 
detection, we mean the direct measurement of the emission (thermal or 
otherwise) from a ``planet'' and its central star, and hence the 
ability to produce an image of the star and the planet, albeit with 
only 1 pixel on each object.  For the purposes of this section, 
``planet'' means a distinct body orbiting a central star.  The only 
such bodies luminous enough to detect are gaseous giant planets.  
However, they may be at any stage of their evolution, i.e., they may 
be very young, and hence very large and hot, or they may be quite 
mature (like Jupiter).  The conditions necessary for such a detection 
are [1.] there must be sufficient flux density from the planet to 
              obtain a ``reasonable'' signal to noise ratio (SNR), in a 
	      ``reasonable'' amount of time; [2] the emission from the planet must be distinguished from 
	      that from the star; and [3.] the detection must be obtained in a short enough time that
	      the planet does not move too far in the plane of the sky.
For observations of any thermal blackbody (with emission which goes
like $\lambda^{-2}$), and given the atmosphere and current detector technology, 
the SNR of a detection is maximized at 345~GHz
(since SNR$\propto {\nu^2 \over {\Delta{S}}}$, where $\Delta{S}$ is the
noise at frequency $\nu$.)

%
%

As examples, we will consider 3 different types of giant planets, 
corresponding to different evolutionary ages.  First, a mature giant 
planet similar to our own Jupiter: $R \sim 7.0 \times 10^7 {\rm m} \sim 
1 R_j $; $T \sim 200$\ K.  Second, a mature, but hotter planet (which 
might be considered a brown dwarf, e.g., Gl229B): $R \sim 1.0 \times 
10^8 {\rm m} \sim 1.5 R_j $; $T \sim 1000$\ K.  And lastly, a very 
young ``protoJupiter'': $R \sim 2.1 \times 10^{9} {\rm m} \sim 30 
R_j $; $T \sim 2500$\ K.  Values of the flux density for these objects 
at 1, 5.7, 10 and 120 parsecs are shown in Table \ref{flux}, for a frequency of 
345 GHz.

\begin{table}[tbh]
\caption{Giant planet flux densities at 345 GHz (in $\mu$Jy). }
\vspace*{1truemm}
\footnotesize
\label{flux}
\begin{center}
\begin{tabular}{cccc}
\hline\hline
\noalign{\vspace{3pt}}
distance (pc) & Case 1 (Jupiter) & Case 2 (Gl229B) & 
   Case 3 (protoJupiter) \\
\noalign{\vspace{3pt}}
\hline
\noalign{\vspace{3pt}} 
1   & 12     & 130   & 59000 \\
5.7 & 0.36   & 4.1   & 1820  \\
10  & 0.12   & 1.3   & 590   \\
60 & 0.004 & 0.043 & 43  \\
120 & 0.0008 & 0.009 & 4.1   \\
\noalign{\vspace{3pt}}
\hline\hline
\end{tabular}
\end{center}
\footnotesize
\end{table}

The detection
of any mature giant planet (like Jupiter) would only be feasible for
the very nearest stars. However, the direct detection of the very
young, hot, protoJupiters in the nearest star forming regions (e.g. T Hya at 56pc) 
could be achieved in six hours at the 5$\sigma$ level.


The separation of the protoJupiter from the central protostar must be
sufficiently large to allow for discrimination between the two.  For a
face-on system, this separation is simply: $\theta_{sep} = a / D$.
Very high resolution is needed, as for e.g. observations at 345 GHz with
maximum baselines of 4 km, the resolution will be $\sim 45$
milliarcseconds.  For a system at 120 pc with a protoJupiter at 5 AU
orbital radius, $\theta_{sep} \sim 40$ milliarcseconds, requiring 15 km baselines offering
resolution of $\sim 10$ milliarcseconds.

The above treatment has ignored the flux density from the central 
protostar.  At optical and infrared wavelengths, the confusion from 
the central protostar may be a problem, as it can be as much as 6 
orders of magnitude brighter than the protoJupiter.  Fortunately, this 
ratio is much less at sub-mm and mm wavelengths.  
Typical ratios at these wavelengths should be only on the order of 1000 
or so, posing no dynamic range problem for interferometers.  In fact, the flux 
density from the central protostar is an aid in imaging with interferometers with sufficient reslolution, 
as it could be used to maintain the coherence of the instrument 
(provided its flux density is large enough).

{\bf Indirect Detection (Astrometry).}
The orbit of any planet around its central star causes that star to 
undergo a reflexive circular motion around the star-planet barycenter.  
By taking advantage of the incredibly high resolution of radio interferometers, 
we may be able to detect this motion.  

The astrometric resolution, or the angular scale over which 
changes can be discriminated ($\Phi$), is proportional to the intrinsic 
resolution of an instrument $ \theta_{HPBW} $ and inversely proportional to the signal to noise 
with which the stellar flux density is detected (${\rm SNR}_*$).
This relationship provides the key to high precision astrometry: the 
astrometric accuracy increases both as the intrinsic resolution 
improves and also as the signal to noise ratio is increased.  Radio emission may be detected from
many stars.  Nonthermal emission has been detected from many varieties of stars, and thermal emission will be easily detected by sufficiently sensitive submillimeter telescopes.  Late-type stars are particularly numerous, bright and nearby enough to form excellent targets for astrometric study by very long baseline techniques and will be more productive as sensitivity of long baseline arrays increases.  Stars of many spectral types will be detectable with sensitive submillimeter arrays.
Astrometry at radio wavelengths routinely achieves absolute astrometric 
resolutions 100 times finer than the intrinsic resolution, and can 
achieve up to 1000 times the intrinsic resolution with special care.  
For an instrument with stringent phase stability specifications such astrometric 
accuracy can be achieved for wide angle astrometry.  

\begin{figure}[!ht]
\begin{center}
\includegraphics[scale=0.3, angle=0]{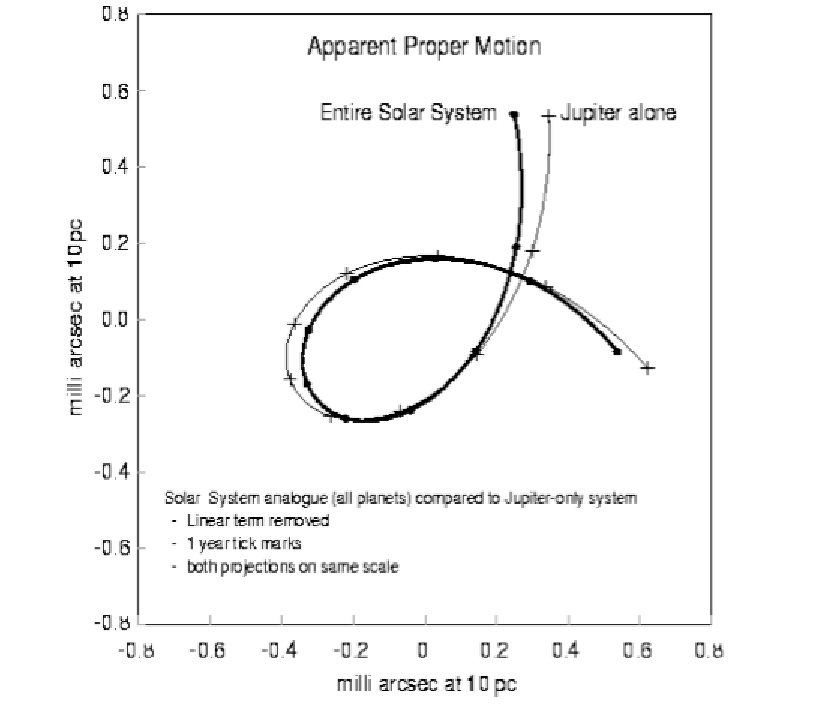}
\caption{Amplitude of the astrometric wobble
 of a solar system analogue over ten years, viewed at ten parsecs.  With short observations spaced several months apart, the motion could be measured.  }\label{fig2}
\end{center}
\end{figure}

When the astrometric resolution is less than the reflexive orbital 
motion, that is, when $\Phi \ \simlt \ \theta_r $, 
that motion may be detected.  


Observations with an interferometer at 345 GHz ($\lambda \sim 0.87$ mm), and baseline lengths
 ($B_{max} \sim 15$ km) 
will detect the reflex motion of 
systems with such planets as far away as ${\rm SNR}_*$ parsecs 
distant (i.e., the ratio of ${\rm SNR}_*$ to $D_{pc}$ is roughly 1 for
our own giant planets).  So, e.g., if we can reach an SNR of 10 on the 
central star, we can detect companions around such stars to 10 parsecs.
If we use the expected thermal flux density of the Sun at 345 GHz as a 
guide, the received flux density from other similar stars 
could achieve an 
${\rm SNR}_*$ of 10 should be achieved in about 10 min for a Sun at 
10 parsecs distance.  Note, however, that astrometric detection of a 
planet requires that curvature in the apparent stellar motion be 
measured, since linear terms in the reflex motion are indistinguishable 
from ordinary stellar proper motion.  This implies that at the very 
minimum, one needs three observations spaced in time over roughly half 
of the orbital period of the observed system.   A detection of a 
planetary system with astrometry would thus require some type of 
periodic monitoring.  

In addition to monitoring of the stellar motion, monitoring of over-densities
(clumps) in debris disks could also provide indirect evidence of an orbiting planet
(e.g. the Vega debris disk shown in Figure 2); as the planet orbits the central star,
planetesimals trapped in resonances will follow the planetÕs motion. This has been
theoretically modeled for the Vega debris disk system (Wyatt (2003)), and tentatively
detected in the $\epsilon$ Eridani debris disk (Poulton, Cameron \& Greaves (2006)).


If all of the detectable stars had planetary companions, how many of
them could be detected (via astrometry) with ALMA?  A study of the Hipparcos and
Gliese catalogs shows that hundreds of stars accompanied by planets of 1 Jovian mass
or more could be measured astrometrically.  Several dozen exist around which a Neptunian
solar mass planet could be detected.



Ground based efforts to find extra solar planets have focused in recent 
years in several areas, including differential astrometry, radial 
velocity measurements, and gravitational lensing.  There are also 
proposals for methods which could directly image Jovian-class planets 
near a small number of stars.  Planet detection using astrometry with 
radio interferometry will both complement these efforts, and have several important 
advantages:  
   $\bullet${Astrometric searches for planets with radio interferometry will use 
         absolute (wide-angle) astrometry, avoiding the problem of 
	 solving for motions in both a target star and a background 
	 reference star.  Stellar positions may be tied 
	 directly to the quasar reference frame, since both stars and 
	 quasars will be bright enough for high precision astrometry . }
   $\bullet${Systematic errors will be lower with radio interferometers than with 
	 techniques at optical or IR wavelengths because they can 
	 observe stars 24 hours a day, providing various types of 
	 closure constraints on their astrometry.  This will reduce the 
	 kinds of seasonal systematic errors that plague optical 
	 astrometry. }
	 $ \bullet${Astrometry measures the position of the star, so that the mass 
	 of the unseen planet can be easily determined.  Direct 
	 detection techniques, either ground or space-based, cannot 
	 determine planetary masses, but will provide constraints on 
	 the parameters of a particular planetary system that 
	 complement those provided by astrometry. }
	 $\bullet${Astrometric results from will directly complement the 
	 results expected from differential astrometry with proposed 
	 ground-based IR interferometers.  Radio interferometers will not face the 
	 constraints on finding suitable background reference stars 
	 that are required by differential IR astrometry, so some 
	 additional stellar systems may be accessible. }
   $\bullet${As with other astrometric techniques, the detectability of a 
	 planet does not depend on the inclination of the planet's 
	 orbital plane around its primary star.  In fact, astrometry 
	 could resolve inclination ambiguities for planets discovered 
	 using radial velocity techniques, if the amplitude of the 
	 astrometric signal is large enough. }
   $\bullet${Astrometric searches are complementary to radial velocity 
         searches in that the former are more sensitive to planets with 
         larger semimajor axes, and the latter are more sensitive to 
         ones with smaller semimajor axes. }

{\bf Summary.}
Multi-wavelength imaging over wide fields will identify regions of planet formation.  These can then be investigated over broad bandwidths to identify the molecular and dust characteristics of the natal disk.  High spatial (hundredths of an arcsecond or better) and spectral resolution ($\sim 1$ km s$^{-1}$) can resolve kinematics of planet-forming disks.  With very high spatial resolution, astrometry (to tenths of milliarcseconds or better with ALMA and/or VLB techniques) can reveal the presence of planets about many nearby stars, through active regions on late type stars or thermal emission from a cross-section of stars of any spectral type.


\end{document}